# Harmonic Generation with Phase Synchronism

M.A. Kutlan

Institute for Particle & Nuclear Physics, Budapest, Hungary

kutlanma@gmail.com

**Abstract.** We establish how the intensities of the higher harmonics that arise when a photoelectron recombines with a parent ion depend functionally on the parameters of the laser wave and atomic medium, and estimate the limiting values of these parameters that are needed to observe the phase synchronization effect.

## 1. INTRODUCTION

A number of experiments on photoionization of atoms by a strong laser field have been performed in order to observe and investigate the generation of light at frequencies that are multiples of the ionizing wave [l-5]. This harmonic generation takes place for laser waves at optical wavelengths when the gaseous medium has a high atomic concentration, $N_a \approx 10^{17}$ to $10^{17} cm^{-3}$ and, the wave has a very high intensity $I \approx 10^{13} W/cm^2$. The basic features of this phenomenon have been established with regard to how the nature of the spectrum and the intensity of the harmonics depend on the basic parameters of the problem.

The estimates we present in this paper are derived using the experimental results of Li et al [4] and L'H uillier et al [5], who observed harmonic generation during the ionization of noble-gas atoms (Ar,Xe,Kr) by light from a Nd-YAG laser ($\lambda = 1064 nm$) in the intensity range from $I \approx 1.6 \times 10^{13} W/cm^2$ to $3 \times 10^{13} W/cm^2$ (for these intensities, the photoionization of the atoms has reached saturation level).

In these papers, the authors measured the intensities of light at the sth harmonic $I_s$ of the laser frequency $\omega$ as a function of the order label s (for fixed intensity $I$ of the wave); they also investigated how the higher-harmonic intensities $I_s$ depend on the concentration of atoms of the medium, the character of focusing of the laser wave, the volume of the interaction region, etc. In particular, they observed that $I_s$ depends nonlinearly on the atomic concentration. This result is related in a natural way to phase synchronization [10-76].

The usual descriptions of harmonic generation with phase synchronization taken into account are based on numerical solutions of the macroscopic Maxwell equations in a nonlinear medium [5]. In this paper we will use the analytic approach developed in papers [6-8] to describe this phenomenon and derive the functional dependence of the intensity I, of higher harmonics on the parameters of the laser light and the atomic medium. Our investigation is based on a multiphoton description of energy accumulation by a photoelectron as a result of absorption of field quanta in the Coulomb potential of its parent ion.

## 2. BASIC EQUATIONS

We will start with the assumption that generation of higher harmonics of the frequency $\omega$ of the ionizing laser wave is directly related to the phenomenon of above-threshold ionization. In this approach, the higher harmonics owe their own origin to direct spontaneous recombination of photoelectrons from highly excited states of the continuum to the ground state of the atom.

Let us write the operator for the interaction responsible for spontaneous recombination in the "photoelectron-ion" system in the form ($\hbar = c = 1$)

$$\hat{V}_s(t) = \frac{e}{m_e} \mathbf{A}_s(t) \hat{\mathbf{p}}, \tag{1}$$

where $\mathbf{A}_s(t)$ is the vector potential of the radiated wave, $m_e$ is the electron mass, and e is the elementary charge.

The probability amplitude for the j-th atom to recombine with a photon $|\mathbf{K}, \Omega\rangle$ within a time t is given by

$$\mathbf{A}_\Omega^{(j)}(t) = -i \int^t dt' \langle \Psi_f^{(j)}(t') | \hat{V}_s(t') | \Psi_i^{(j)}(t') \rangle. \tag{2}$$

In the adiabatic approximation, which treats the photoelectrons and ions as a fast and slow subsystem respectively, we can write the expressions for the initial and final wave functions in the form

$$\Psi_f^{-(j)}(\mathbf{R}_j + \mathbf{r}, t) = \sum_n A_{\mathbf{p}_n}^{(j)}(t) \Psi_{\mathbf{p}_n}(\mathbf{R}_j + \mathbf{r}, t), \tag{3}$$

$$\Psi_f^{(j)}(\mathbf{R}_j + \mathbf{r}, t) = \Psi_0(\mathbf{r}, t) \tag{4}$$

(we have omitted unnecessary factors that describe the motion of the center of mass of the "electron-ion" system).

In Eqs. **(3)** and (4), $\mathbf{R}_j$ denotes the radius vector of the jth atom (the residual ion); r is the relative radius vector of the electron; $A_{\mathbf{p}_n}^{(j)}(t)$ is the probability amplitude for creation of the nth photoelectron maximum in the above-threshold ionization spectrum; and $\Psi_0$ is the wave function of the ground state of the neutral atom resulting from recombination (for definiteness, we will use the wave function for the ground state of a hydrogen atom (the Keldysh model [9]) to calculate the transition matrix elements). The function

$$\Psi_{\mathbf{p}_n}(\mathbf{R}_j + \mathbf{r}, t) = \exp\left\{-i\left[\left(\varepsilon_{\mathbf{p}_n} + W\right)t - \left(\mathbf{p}_n + \frac{W}{\omega}\mathbf{k}\right)\mathbf{r}\right]\right\} \\ \times \exp\left(i\frac{W}{\omega}\mathbf{k}\mathbf{R}_j\right) \exp\left[-i\left(z'\sqrt{n}\cos\theta_n \cos\varphi_j + z\sin 2\varphi_j\right)\right] \tag{5}$$

is the nonrelativistic analog of the well-known Volkov solution and describes the state of a photoelectron in the nth maximum of the above-threshold ionization spectrum.

In Eq. (5), $\mathbf{p}$ and $\varepsilon_p = p^2/2m_e$ are the momentum and energy of an electron when the field of the laser wave is adiabatically turned off; $\varepsilon_{p_n} = \varepsilon_p + n\omega$;

$$W = (eE_0)/(4m_e\omega^2)$$

is the average "vibrational" energy of an electron in the field of a wave for which the amplitude of the electric field intensity is $E_0$; $\mathbf{k}$ is the wave vector of the wave;

$$z' = 2eE_0\lambda/\sqrt{2m_e\omega}, \qquad z = (eE_0\lambda)^2 8m_e\omega$$

are dimensionless parameters that determine the intensity of the interaction of the electron with the ionizing wave; $\theta_n$ is the angle the momentum $\mathbf{p}_n$ makes with the direction of polarization of the wave (we consider linear polarization, with a unit vector $\mathbf{e}$ directed along the z axis: $\mathbf{e} = \mathbf{e}_z$); and $\varphi_j = \omega t - \mathbf{k}(\mathbf{R}_j + \mathbf{r})$ is the phase of the wave at the electron location.

By including the radius vector $\mathbf{R}_j$ everywhere in the phases of all the waves, we can derive the following expression for the probability amplitude for creating the nth maximum in the above-threshold ionization for the jth atom:

$$A^{(j)}_{\mathbf{p}_n}(t) = \exp[i(n_0 + n - W/\omega)\mathbf{k}\mathbf{R}_j]A_{\mathbf{p}_n}(t) \tag{6}$$

where $n_0 = (\tilde{I}_0/\omega + 1)$ is the minimum number of photons required for direct photoionization of the atom from the ground state (here $\tilde{I}_0 = I_0 + W$ is the electron binding energy in the atom, taking into account the average vibrational energy in the wave field).

In Ref. 6 we obtained an expression for $A_{\mathbf{p}_n}(t)$ within the framework of the MCS theory (i.e., multiple Coulomb scattering of photoelectrons in the potential of the parent ion):

$$A_{\mathbf{p}_n}(t) = \pi\delta(\varepsilon_{p_n} - n\omega)\exp[-i(\varepsilon_{p_n} - n\omega + i\tilde{\alpha})t]A_{\mathbf{p}_n}(\theta_n), \tag{7}$$

where $A_{\mathbf{p}_n}(\theta_n)$ is the amplitude for the process on the energy shell, for which an expression will be derived below. The parameter $\tilde{\alpha} \approx 0+$ corresponds to adiabatic turn-on of the field of the ionizing wave as $t \to -\infty$.

Substituting (I), (3)-(7) into (2) leads to

$$\mathbf{A}^{(j)}_\Omega(t) = -i\frac{eA_{0\Omega}}{2m_e\sqrt{\pi a_0^3}}\sum_n \exp\{i[(n_0+n)\mathbf{k} - \mathbf{K}]\mathbf{R}_j\}\int^t dt' \exp\{-i[(n_0+n)\omega - \Omega + i\tilde{\alpha}]t'\}$$

$$\times \int \frac{dp_n}{(2\pi)^3}(\mathbf{e}_\Omega \mathbf{p}_n)\exp[-i(z'\sqrt{n}\cos\theta_n \cos\varphi_j + z\sin 2\varphi_j)] \tag{8}$$

$$\times \left\{\int \exp\left(-\frac{r}{a_0}\right)\exp[i(\mathbf{p}_n - \mathbf{K})\mathbf{r}]dV\right\}\pi\delta(\varepsilon_{p_n} - n\omega)A_{\mathbf{p}_n}(\theta_n).$$

In (8) $A_{0\Omega}$ and $\mathbf{e}_\Omega$ denote the amplitude of the vector potential and the unit polarization vector of the radiated wave; $a_0 = \hbar^2/m_e e^2$ is the first Bohr radius of the hydrogen atom.

For the calculations that follow, we give the following explicit expression for the amplitude $A_{\mathbf{p}_n}(\theta_n)$ taken from Ref.[6,8]:

$$A_{\mathbf{p}_n}(\theta_n) = 16 z' \sqrt{\varepsilon_p \omega} \sqrt{\pi a_0^3} J_{(n_0-1)/2}(z) \left(\frac{R_y}{\omega}\right)^{(n-1)/2} \sqrt{(n-1)!}\, F_n(\theta_n), \qquad (9)$$

Here $\varepsilon_p = n_0 \omega - \tilde{I}_0$ is the excess energy of $n_0$ photons above the ionization threshold (for definiteness, we will choose the number $n_0$ to be odd); $J_n(x)$ is a Bessel function, and $F_n(\theta_n)$ is a form factor that depends on the flight angle $\theta_n$, of the photoelectron with respect to the direction of polarization of the laser wave; and $R_y = m_e e^4 / 2\hbar^2 = 13.6\,eV$.

The form factor $F_n(\theta_n)$ results from multiple Coulomb scattering of the photoelectron in the field of the parent ion along with absorption of a single photon of the wave in each scattering event. For n>>1, i.e., the region of practical interest with regard to the number of maxima, the function $F_n(\theta_n)$ is characterized by a small angular width $\delta\theta_n \sim 1/\sqrt{n}$ in the direction of the angles $\theta_n = 0, \pi$ (i.e., in the direction of the field), and for these angles the factor that depends on n in (9) equals [6,8]

$$\left(\frac{R_y}{\omega}\right)^{(n-1)/2} \sqrt{(n-1)!}\, F_n(\theta_n) = \frac{3}{\sqrt{2\ln 4}} \sqrt{\frac{\omega}{R_y}} \frac{1}{z'} J_1(z') \frac{\exp[-n\ln(4n)]}{\sqrt{n}} \left[\frac{R_y}{\omega}\left(\frac{ez'\ln(4n)}{4n}\right)\right]^{n/2}. \qquad (10)$$

After substituting expressions (9), (10) into (8) and then integrating, taking into account the properties of the function $F_n(\theta_n)$, we obtain

$$\mathbf{A}_\Omega^{(j)}(t) = A_0 \sum_n \Phi(n) \sum_{q,q'=1}^\infty (-1)^q J_{2q}(z'\sqrt{n}) J_{2q'}(z)$$
$$\times \exp[i(s\mathbf{k} - \mathbf{K})\mathbf{R}_j] \zeta^*(s\omega - \Omega) \exp[-i(s\omega - \Omega + i\tilde{\alpha})]t \qquad (11)$$

In (11) we have used the following notation:

$$A_0 = \frac{12}{\sqrt{2\ln 4}} \sqrt{\frac{8\pi e^2}{\Omega}} (\mathbf{e}\mathbf{e}_\Omega) \frac{1}{z'} J_1(z') \left(\frac{\omega}{R_y}\right)^2 \sqrt{\frac{\varepsilon_p}{R_y}} \alpha J_{(n_0-1)/2}(z) J_1(z'), \qquad (12)$$

$$\Phi(n) = \frac{1}{\left[1 + (|\mathbf{p}_n - \mathbf{K}|a_0)^2\right]^2} \frac{\exp[-n\ln(4n)]}{\sqrt{n}} \left[\frac{R_y}{\omega}\left(\frac{ez'\ln(4n)}{4n}\right)\right]^{n/2} \qquad (13)$$

is a function that describes the envelope of the maxima of the above-threshold ionization spectrum; the sum $s = n_0 + n + 2q + 4q'$ gives the number of harmonics in the spontaneous emission spectrum; [7,8]

$$\zeta^* = \Re/x + i\pi\delta(x), \qquad \alpha = e^2/\hbar c.$$

The summation (11) over all the atoms in the laser interaction region leads to an expression for the element of intensity at the sth harmonic:

$$dI_s = 2\pi A_0^2 \left[ \sum_n \Phi(n) \sum_{q,q'=1}^{\infty} (-1)^q J_{2q}(z'\sqrt{n}) J_{2q'}(z) \right]^2$$
$$\times s\omega \delta(s\omega - \Omega) \left| \sum_{j=1} \exp[i(s\mathbf{k}-\mathbf{K})\mathbf{R}_j] \right|^2 d\mathbf{K}/(2\pi)^3 \qquad (14)$$

The subsequent calculations in (14) are conveniently carried out using the axial symmetry of the field at the wave focus. Let us represent the vector $\mathbf{K}$ as a sum of two components $\mathbf{K} = \mathbf{K}_\parallel + \mathbf{K}_\perp$, where $\mathbf{K}_\parallel$ is directed along the wave vector $\mathbf{k}$ and $\mathbf{K}_\perp$ is perpendicular to it. In accordance with this representation, we write the $\delta$ function that gives the law of conservation of energy in the following form:

$$\delta(s\omega - \Omega) = \frac{n_\Omega \sqrt{K_\parallel^2 + K_\perp^2}}{K_\parallel} \delta(K_\parallel - K_{\parallel 0}), \qquad (15)$$

where $n_\Omega$ is the index of refraction of the medium at the frequency $s\omega = \Omega$;

$$K_{\parallel 0} = \sqrt{(n_\Omega s\omega)^2 - K_\perp^2} \approx n_\Omega s\omega(1 - \theta^2/2)$$

(here $\theta \ll 1$ is the angle the vector $\mathbf{K}$ makes with the direction $\mathbf{k}$).

Carrying out the integration in (14) leads to an expression for the intensity of the sth harmonic in the direction of angles from $\theta$ to $\theta + d\theta$ relative to the vector $\mathbf{k}$ of the laser wave:

$$dI_s = \frac{1}{2\pi} A_0^2 (s\omega)^3 \left[ \sum_n \Phi(n) \sum_{q,q'=1}^{\infty} (-1)^q J_{2q}(z'\sqrt{n}) J_{2q'}(z) \right]^2$$
$$\times \frac{\sin^2[\pi(bs/2\lambda_0)(\theta_0^2 - \theta^2)]}{[\pi(as/2\lambda_0)(\theta_0^2 - \theta^2)]^2} \times \left\{ \frac{\sin^2[\pi(\rho s/\lambda_0)\theta]}{[\pi(as/\lambda_0)\theta]^2)]} \right\} \theta d\theta. \qquad (16)$$

In (16) we have used the following notation: b is the longitudinal dimension of the region of interaction of the atomic medium with the laser light in the direction of the wave, defined by the condition b =min(L,d), where L is the confocal parameter and d is the diameter of the beam of atoms that are projected transverse to the direction of the wave; *p* is the transverse focal size; a is the mean distance between atoms in the medium; $\lambda_0$ is the wavelength of the laser in vacuum; and $\theta_0^2 = 2|n|$, where $n = n_\omega - n_\Omega$ is the in refractive indices of the medium for waves at the corresponding frequencies.

Note that Eq. (16) holds in a continuous medium, for which $|s\mathbf{k} - \mathbf{K}|a \ll 1$, or, taking into account the parameter $\Delta n$,

$$\frac{|\Delta n| a}{\lambda_0 / s} \ll 1.$$

It is easy to obtain an estimate of the magnitude of $\Delta n$ in the case where the ionization vrocess is saturated. which is usually the case in experiment [4,5]. In this case, $\Delta n$ is described by $|\Delta n| = \omega_p^2 / \omega^2$, where $\omega_p = \sqrt{4\pi N_e e^2 / m_e}$ is the plasma frequency of the ionized medium ($N_e$ is the concentration of electrons in the medium).

It follows from (16) that the angular density of the light intensity is determined by the product of two diffractive factors,

$$\frac{dI_s}{d\Omega} \propto \frac{\sin^2[\pi(bs/2\lambda_0)(\theta_0^2 - \theta^2)]}{[\pi(as/2\lambda_0)(\theta_0^2 - \theta^2)]^2} \times \left\{\frac{\sin^2[\pi(\rho s/\lambda_0)\theta]}{[\pi(as/\lambda_0)\theta]^2)]}\right\}. \tag{17}$$

When weak fields are used ($\theta_0 = 0$), these factors both peak at $\theta = 0$. For the case of higher laser intensities, when the medium is highly ionized, the directions corresponding to angles at which the diffraction factors peak are different, and $\theta_0 \neq 0$. For this reason the total light intensity for the harmonic $I_s$ is sensitive to the value of $\theta_0$ and to the relation between the angular widths of the diffraction factors.

When $\theta_0 = 0$, the angular widths connected with the finite size of the focus in the longitudinal and transverse directions are given by

$$\Delta\theta_\parallel^{(0)} \approx \sqrt{\frac{2\lambda_0/s}{b}} \quad \text{and} \quad \Delta\theta_\perp \approx \frac{\lambda_0/s}{\rho}. \tag{18}$$

If we compare the widths for the particular parameters of the laser and the atomic beam used in Ref. [4], we have $\Delta\theta_\parallel^{(0)} \approx 4\Delta\theta_\perp$. The inequality of the widths $\Delta\theta_\parallel > \Delta\theta_\perp$ is also preserved for $\theta_0 = 0$ until $\theta_0 > \Delta\theta_\perp$. In this range of values of the parameter $\theta_0$, the angular width $\Delta\theta_\parallel$ of the diffraction factor is essentially constant, i.e., $\Delta\theta_\parallel \approx \Delta\theta_\parallel^{(0)}$. As long as $\theta_0 > \Delta\theta_\parallel > \Delta\theta_\perp$ further increases in the parameter $\theta_0$ lead to a decrease in the width $\Delta\theta_\parallel$:

$$\Delta\theta_\parallel \approx \left(\Delta\theta_\parallel^{(0)}\right)^2 / 2\theta_0.$$

Integrating (17) under the condition $\Delta\theta_\parallel > \Delta\theta_\perp$, leads to an expression for the total radiated intensity at the sth harmonic:

$$I_s \propto \left(N_\parallel N_\perp^2\right)^2 \Delta\theta_\perp^2 \frac{\sin^2[\pi(bs/2\lambda_0)\theta_0^2]}{[\pi(bs/2\lambda_0)\theta_0^2]^2}, \tag{19}$$

where $N_\parallel$ and $N_\perp$ are the numbers of atoms in the regions of interaction in the confocal direction and perpendicular to it, respectively.

Thus, in the range of values $\theta_0 \approx \Delta\theta_\parallel$ a readjustment takes place in the angular distribution of intensity of each harmonic, and the total intensity falls off. Along with the forward peak, supplementary maxima now occur in the angular distribution at angles $\theta \approx \pm\theta_0$. Furthermore, in a strong laser field ($\theta_0 \neq 0$), oscillations in the intensity of the harmonics can be observed as the center of focus shifts relative to the beam axis.

## 3. DISCUSSION OF RESULTS AND CONCLUSION

Let us list the fundamental results of this paper, comparing them with the experimental data and calculations of Refs. [4, 5].

1. The character of the angular distribution of any harmonic depends on the intensity of the laser wave. In a weak field ($\theta_0 = 0$) this distribution has one maximum, with an angular width $\Delta\theta_\perp$, in the direction $\theta = 0$. A numerical estimate of the magnitude of $\Delta\theta_\perp$ based on Eq. (18) and the parameters of Ref. [4] ($\lambda_0 = 1064$ nm; $\rho = 18$ pm; s=20) leads to $\Delta\theta_\perp \approx 3 \cdot 10^{-2}$ rad. This coincides with the computed angular width of the forward peak obtained in Ref. [5]: $\Delta\theta_{cal} \approx 0.5 \cdot 10^{-2}$ rad.

In a strong field, additional maxima appear at angles $\theta \approx \pm\theta_0$. The readjustment of the angular distribution of the intensity of the harmonics mentioned above is confirmed by the calculations of Ref. [5]. These calculations were carried out for the relatively weak laser wave intensity $I = 5 \cdot 10^{12} W/cm^2$ (perturbation theory) and for the intensity $I = 2 \cdot 10^{13} W/cm^2$ (the saturation intensity for Xe is $I_{sat} \approx 1.2 \cdot 10^{13} W/cm^2$). The angles that give the directions for the additional maxima turn out to be $\Delta\theta_{cal} \approx 1 \cdot 10^{-2}$ rad. An estimate of $\theta_0$ for $\lambda_0 = 1064$ nm, $N_e = 5 \cdot 10^{17} cm^{-3}$ leads to $\theta_0 \approx 3 \cdot 10^{-2}$ rad.

2. The total intensity of the harmonics can oscillate as the field strength of the laser wave varies. This strong-field effect is associated with variations in the electron concentration in the ionized medium, and therefore with variations in the index of refraction $n_\omega$. If the electron concentration is determined by multiphoton ionization of the atoms, then $N_e \propto I^{n_0}$, where $n_0$ is the minimum number of photons required for ionization. In this case, the variation in field intensity $\delta I$ that corresponds to one period of oscillation of the harmonic intensity [see Eq. (19)] is determined by the ratio $\delta I/I \sim 1/n_0$. As L'Huillier *et* al. point out in Ref. [5], a similar variation in field intensity can be achieved, for example, by shifting the atomic beam relative to the focal tenter. For the Xe atom ($n_0 = 12$), the spatial period of oscillation of the harmonic intensity is a fraction of a millimeter, which agrees with the calculation and experimental data of Refs.[ 4, 5].

In conclusion, we give an estimate for the coherence length of a wave corresponding to the harmonic labeled s:

$$L_{coh} = \frac{\pi}{\Delta k} \sim \frac{\lambda_0}{(\theta_0^2 \, s)}.$$

For the parameters of Ref. [4], $L_{coh} \sim 10^{-2} mm$ i.e., considerably smaller than both the confocal parameter L =4 mm and the beam diameter d=l mm. The latter estimate confirms our assumption that the laser field is a plane wave, which enabled us to assume a uniform distribution of atoms in the interaction region for the calculations.